\documentclass[a4paper,twoside,10pt]{letter}
\usepackage{graphicx,saj,multicol,subeqnarray}

\def\papertype{}
\newcommand{\D}{$^\circ$}

\begin{document}
\baselineskip=3.1truemm
\columnsep=.5truecm
\newenvironment{lefteqnarray}{\arraycolsep=0pt\begin{eqnarray}}
{\end{eqnarray}\protect\aftergroup\ignorespaces}
\newenvironment{lefteqnarray*}{\arraycolsep=0pt\begin{eqnarray*}}
{\end{eqnarray*}\protect\aftergroup\ignorespaces}
\newenvironment{leftsubeqnarray}{\arraycolsep=0pt\begin{subeqnarray}}
{\end{subeqnarray}\protect\aftergroup\ignorespaces}
%


\markboth{\eightrm
Evidence of molecular adaptation to extreme environments and
applicability to space environments}
    {\eightrm M. FILIPOVI{\' C},
S. OGNJANOVI{\' C} and M. OGNJANOVI{\' C}}

{\ }

\publ

\type

{\ }


\title{
Evidence of molecular adaptation to extreme environments and
applicability to space environments}


\authors{M. Filipovi{\' c}$^{1}$, S. Ognjanovi{\' c}$^{2}$ and M. Ognjanovi{\' c}$^{2}$}

\vskip3mm


\address{$^1$University of Western Sydney, Locked Bag 1797, Penrith South DC,
    NSW 1797, Australia}

\address{$^2$University of Minnesota, MMC 715, 1-185 Moos 420 Delaware Street SE Minneapolis, MN
55455, USA}


\dates{20 February, 2008}{1 April, 2008}


\summary{This is initial investigation of a gene signatures
responsible for adapting microscopic life to the life in extreme
Earth environments. Here, we present a preliminary results on
identification of the clusters of orthologous groups (COGs) common
to several hyperthermophiles and exclusion of those common to a
mesophile (non-hyperthermophile): {\it Escherichia coli (E. coli K12)}, 
will yield a group of proteins possibly involved in adaptation to life 
under extreme temperatures. Comparative genome analyses represent a 
powerful tool in discovery of novel genes responsible for adaptation 
to specific extreme environments. Methanogens stand out as the only 
group of organisms that have species capable of growth at 0\D C 
({\it Metarhizium frigidum (M.~frigidum)} and {\it Methanococcoides 
burtonii (M.~burtonii)}) and 110\D C ({\it Methanopyrus kandleri 
(M.~kandleri)}). Although, not all the components of heat adaptation 
can be attributed to novel genes, the {\it chaperones} known as heat 
shock proteins stabilize the enzymes under elevated temperature. 
However, highly conserved {\it chaperons} found in
bacteria and eukaryots are not present in hyperthermophilic Archea,
rather, they have a unique {\it chaperone TF55}. 
Our aim was to use software which we specifically developed for 
extremophile genome comparative analyses in order to search for additional 
novel genes involved in hyperthermophile adaptation. The following
hyperthermophile genomes incorporated in this software were used for
these studies: {\it Methanocaldococcus jannaschii (M.~jannaschii), 
M.~kandleri, Archaeoglobus fulgidus (A.~fulgidus)} and three
species of {\it Pyrococcus}. Common genes were annotated and grouped
according to their roles in cellular processes when such information
was available and proteins not previously implicated in the
heat-adaptation of hyperthermophiles were identified. Additional
experimental data is needed in order to learn more about these
proteins. To address a non-gene based components of thermal
adaptation, all sequenced extremophiles were analysed for their GC
contents and aminoacid hydrophobicity. Finally, we develop a prediction 
model for optimal growth temperature.}



\begin{multicols}{2}
{


\section{1. INTRODUCTION}

Understanding life in extreme environments on Earth can tell us a
great deal about the potential for life in similar environments on
other celestial object such as planets, satellites, comets and
asteroids. Understanding the limits of life as we know can also help
determine what makes a planet habitable.

Astrobiology has developed as a new field, devoted to the scientific
study of life in the universe -- its origin, distribution, evolution
and future. This multidisciplinary field brings together the
physical and biological sciences to address some of the most
fundamental questions of the natural world: the origin of life,
evolution of habitable worlds and adaptations of terrestrial life
required for potential survival beyond our home planet.

We now realise that the origin and evolution of life itself cannot
be fully understood unless viewed from a larger perspective than
just our own planet -- Earth. Biologists are working with
astronomers to describe the formation of life's biochemical
precursors, and to discover new potentially habitable planets, while
collaborations with computer scientist, geologists, paleontologists,
evolutionary biologists, climatologists and planetary scientists
help studies of other aspects of life limits.

Our intention is to investigate a gene signature responsible for
adapting microscopic life forms to the life in extreme Earth
environments with the goals to:
\item{(i)} test if computationally identified genes are expressed in a group of
psychrophiles
\item{(ii)} characterize computationally identified proteins and their functions
\item{(iii)} to investigate whether such genes can be used to modify organisms
which can be used for terraforming suitable planets.
\item{(iv)} predict what is the range of environmental conditions on the other
planets and solar bodies that allows the existence of the basic life
forms

Revelations about extremophiles have invigorated the field of
astrobiology (Feller \& Gerday 2003). In recent years, the field has
also been stimulated by the discoveries of apparently biogenically
derived methane on Mars (Onstott et al. 2006), the knowledge that
methanogens exist and are active in the cold, and that methanogens
can grow and metabolize in Martian-soil stimulant (Cavicchioli
2002). Other exciting findings have been the discoveries of live
microorganisms in ice cores taken from sea ice (Price 2007; Tung
2005) and the presence of water in cold environments on Mars (ice
sheets and permafrost), Mercury and Europa (sediments deep beneath
the icy crust (Christner et al. 2001)). Cryopreserved
micro-organisms can remain viable (in a deep anabiotic state) for
millions of years frozen in permafrost and ice. {\it Psychrophiles}, cold
loving bacteria, proliferate at temperatures 0\D--10\D C, metabolize
in snow at ice at --20\D C, are predicted to metabolize at --40\D C
and can survive at --45\D C (Goodchild et al. 2004; Feller \& Gerday
2003; Margesin 1999; Price \& Sowers 2004; Sounders et al. 2003;
Siddiqui \& Cavicchioli 2006; Wagner et al. 2005). It is estimated
that more than 80\% of biosphere is permanently below 5\D C
(Cavicchioli et al. 2000).

All components of {\it psychrophiles}, must be adapted to cold to enable
an overall level of cellular function that is sufficient for growth
and survival. Cold adaptations in bacteria affects most structural
and functional components of the cell; ranging from the outer
membranes (lipid composition) to the inner cellular machines
(ribosomes), protein translation processes, enzymes and nucleic
acids (tRNA) (Feller \& Gerday 2003; Margesin \& Schinner 1999).
Often there are other life-limiting factors present in these cold
environments, such as high pressure (deep sea), high levels of UV
irradiation (snow and ice cap communities), aridity (Antarctic
cryptoendoliths), low light (cave-dwelling). Despite some advances
in understanding molecular adaptations to cold (D'Amico, et al.
2002; Demming 2002; Feller \& Gerday 2003), these adaptations remain
poorly understood. The studies have shown that psychrophilic
metabolic activities may contribute to weathering processes and
carbon/nutrient cycling (Skidmore et al. 2000; 2005; Hearn
2003) and that these organisms may be utilized for biotechnological,
agricultural, industrial purposes, as well as potential
bioremediation in cold regions (Cavicchioli 2002).

\section{2. AIMS AND APPROACH}

A recent study of NASA Ames Institute on Atacama Desert (the driest
desert on Earth) showed that life on this planet is limited by the
presence of water (Navaro-Gonzales et al. 2003). Understanding the
limits of life on this planet as well as specific adaptations
required for survival in extreme environments represents an
important contribution to the efforts of searching for life or
sustaining it in space.

Our aims are to elucidate the genetic mechanisms underlying the
adaptations to specific extreme environments and the effect of two
physical parameters (pressure and temperature) on adaptation and
limitation of life.

The choice of extremophiles to be studied is based on the hypothesis
that there is water under thick layers of ice on Mars, Mercury and
Europa, concluding that such a water environment would be under high
hydrostatic pressure, high temperature (at the places of hydrothermal 
vents) or low temperature, therefore we chose to study piezophiles, 
hyperthermophiles and psychrophiles, while extreme aquatic habitats 
of hydrothermal vents of Lohihi volcano (Hawaii, USA) and deep
ocean will serve as earth analogues of such space environments.

We plan to develop a database of environments found to date in
space, as well as computer models of hypothetical environments that
could exist in space. We would then create a computer application
interfacing between the extremophile properties and the modeled and
existing environments in order to investigate what kind of organisms
can be expected or cultivated on other planets.

\section{3. METHOD}

Understanding constrains on microbial populations in extreme
environments is of great interest in the context of Earth analogs
for possible extraterrestrial habitats. The application of DNA
microarray technology to studies of life in extreme environments
offers an outstanding opportunity for discovering specific
adaptation to these environments by detecting genes that are
uniquely expressed in the natural environment (specific extremophile 
gene signature).

The evidence of existence of such a specific gene signature is being
gathered on a gene-by-gene basis. For example, {\it Shewanella} genus is
split into 2 major subgenuses: mesophilic pressure-sensitive species
and high pressure-cold-adapted species, the latter shown to produce
large amounts of eicosapentaenoic acid (Kato \& Nogi 2001), which
affects the membrane fluidity, shown to be an important component of
pressure adaptation (MacDonald 1987). Microarray approach proposed
here gives a more global perspective into a great number of genes
affected by a change of a single parameter (such as pressure or
temperature) and yield a better understanding of the changes
required for specific adaptations of microorganisms living under
such conditions (piezophiles, hyperthermophiles). A well defined
hyperthermophile {\it M.~jannaschii}, which genome has been sequenced 
(Bult et al. 1996) was chosen as a starting point in the series of 
the proposed experiments.

Additional insight into specific adaptation of hyperthermophiles
could come through computer analysis of thus far sequenced genomes
of hyperthermophiles and discovery of their gene signature. The
necessity of such an approach is recognized throughout the field (Ng
et al. 2000).

Moreover, the comparison of the genes of all available extremophile
genomes sequences may reveal a group of common genes across these
extremophile microorganisms called here general extremophile gene
signature. Although the majority of extremophiles are confined to a
specific extreme environment, some of them can thrive in more than a
single extreme environment. The latter opens the possibility of
existence of a general extremophile gene signature. An example
supporting this view is Chroococcidiopsis species with its
remarkable tolerance of environmental extremes: forms belonging to
this species are present in a wide range of extreme environments:
from Antarctic rocks to thermal springs and hypersaline habitats
(Friedman \& Ocampo-Friedman 1995).

Genetic engineering has been well established for cyanobacteria and
the methods for insertion of clusters of genes developed (Billi et
al. 2001). This opens a technical possibility of inserting a subset
of genes of interest, for example pressure adaptation genes, into a
mesophile (such as a cyanobacterium species) and testing their
importance in survival under increased hydrostatic pressure. The
concepts of the climate modeling (Meadows et al. 2001) will be
applied towards development of hypothetical models of environments
in space.

\section{4. PRELIMINARY RESULTS}

Our approach of extremophile genome comparisons was first developed
for hyperthermophilic microorganisms, organisms which grow at 90\D C
or higher and have the highest growth temperatures known for life.

We have developed an initial software which incorporated features of
Basic Local Alignment Search Tool (BLAST)\footnote{for more details see: 
http://www.ncbi.nlm.nih.gov/blast} genome and BLAST protein 
and existing databases for several sequenced hyperthermophiles and 
were analysed using COGs (clusters of orthology) as a bases for 
comparisons.

The first step was exclusion of all genes present in bacteria which
do not live in extreme enviroments. We used {\it E.~coli~K12}, a common
laboratory strand, for these purposes. Therefore, the first step of
comparisons was between {\it Pyrococcus abyssi (P.~Abyssi)} (one of the 7
hyperthermophiles analysed) and {\it E.~coli}, which eliminated all the
common COGs and only remaining COGs (around 400) were used for
search of the COGs common to hyperthermophiles.

Our analyses focused on 7 hyperthermophiles with well defined COGs
available in the public domain. The 6 archaeal genomes were chosen
to represent a wide variety of hyperthermal habitats. This approach
significantly reduced the number of common genes which would be
found among the Archaea more closely related, since our goal was the
search for the minimal common genes to all hyperthermophiles. This
goal was limited by the number of currently sequenced hyperthermophile 
genomes and the annotations of those which were sequenced.

\vspace{0.25cm}
\centerline{\includegraphics[width=.47\textwidth]{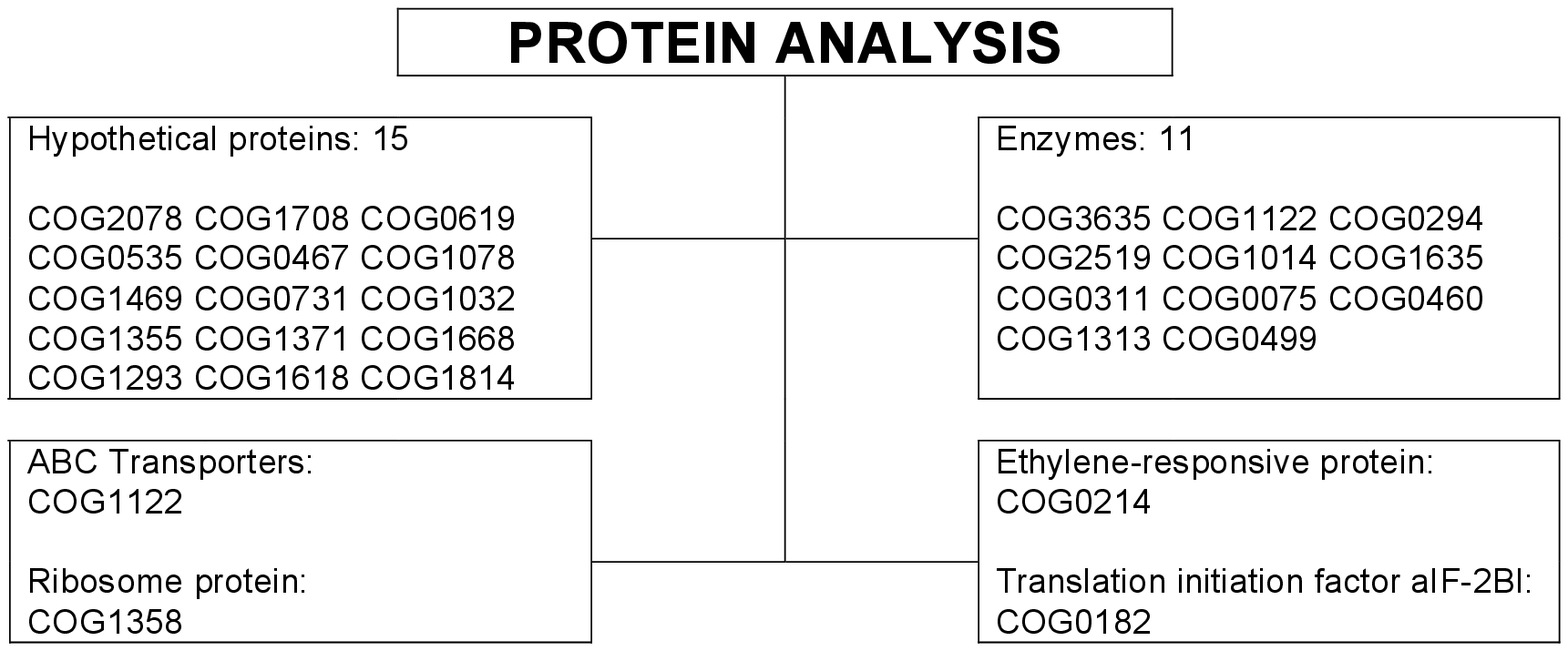}}
\figurecaption{1.}{Hyperthermophile-specific protein groups. Proteins 
were grouped according to function of hypothetical proteins, enzymes, 
sensing systems (ABC transporters) and ethylene-responsive element and 
proteins involved in translation.}

The purpose of these analyses was the identification of genes
involved in adaptation to extreme environment, in this case extreme
temperatures (over 90\D C) and extreme pressure. We identified 29
proteins common to all 7 hyperthermophiles which we grouped
according to function into 4 categories (Fig.~1): hypothetical 
proteins (15 identified proteins had unknown function and were 
annotated as hypothetical), enzymes (11; see Table~1), membrane and 
ribosomal proteins (2) and regulators of gene expression (2).

}

\end{multicols}



{\bf Table 1.} Enzymes common to 7 analysed hyperthermophiles.
Enzymes include a rare enzyme which utilizes Tungsten as a co-factor: 
{\it COG0535}. There have been only 4 such enzymes discovered so far, 
but their importance in a tungsten-reach environment of deep-ocean 
hydrothermal vents is becoming more appreciated.
\vskip2mm
\centerline{\begin{tabular}{|ll|}
\hline
        & Enzymes \\
\hline
{\it COG3635} & Phosphonopyruvate decarboxylase, putative \\
{\it COG0294} & DIHYDROPTEROATE SYNTHASE \\
{\it COG2519} & PROTEIN-L-ISOASPARTATE METHYLTRANSFERASE HOMOLOG \\
{\it COG1014} & 2-ketoglutarate ferredoxinoxidoreductase, subunit gamma (korG-1) \\
{\it COG0311} & Imidazoleglycerol-phosphate synthase, subunit H, putative \\
{\it COG0075} & SERINE-GLYOXYLATE AMINOTRANSFERASE related (EC2.6.1.45) \\
              & (SERINE-GLYOXYLATE AMINOTRANSFERASE)\\
{\it COG0460} & Homoserinedehydrogenase(hom) \\
{\it COG0467} & RECOMBINASE related \\
{\it COG0535} & Tungsten-containing aldehyde ferredoxin oxidoreductase \\
              & COFACTOR MODIFYING PROTEIN \\
{\it COG1313} & Pyruvate formate-lyase activating enzyme (pflX) \\
{\it COG0499} & Adenosylhomocysteinase(ahcY) \\
\hline
\end{tabular}}

\vskip.5cm

\begin{multicols}{2}
{{\ }

At such extreme heat, as found in these environments, proteins are 
expected to loose their tertiary structure, denature, due to coagulation. 
The other challenge is sustaining plasma membrane in semi-liquid state. 
Yet another is the functioning of enzymes at these temperature. The 
categorizes of proteins we identified reflect these issues.

\section{5. DISCUSSION AND FUTURE WORK}

Comparative genomic analyses has the potential to generate
hypotheses regarding the importance of specific genes and molecular
characteristics for life in extremely cold environment, such as the
permafrost (Ponder et al. 2003). The first psychrophile genome
sequenced was {\it Psychrobacter strain 273-4}. It contains a 2.64~Mbp
genome with 2, 147~ORFs. The approach of identification of cold
adaptation genes involved comparison of 2 psychrophile genomes
({\it Psychrobacter 273-4} and {\it Exiguobacterium 255-15}) with the strains of
these bacteria that live in warm waters and can grow at temperatures
up to 42\D C (Ponder et al. 2003). Using this approach, Ponder et
al. (2003) identified one extremely large hypothetical protein
(6,715 amino acids) and four histone-like proteins potentially
involved in cold adaptation. However, comparing the genomes of 2
psychrophiles sequenced at that time, showed 75\% of ORFs in
Exiguobacterium encode for putative protein homologues in
Psychrobacter. Our approach is taking advantage of a larger number
of available sequenced genomes and will use comparisons of these
genomes only and subtraction of genes found in a mesophile ({\it E.~coli
strain~K12}), a method proven successful in hyperthermophile
computational analyses (Sec.~4).

In the Solar system the variety of different conditions exists.
These conditions depend on the distance of the objects from the Sun
and surrounding planet(s) or chemical and physical characteristics
of the object. Therefore, models designed to address possibility of
existence of living forms in our solar system have to include and
understand all factors involved.

From the analyses of comets (Meech et al. 2005; Jones et al. 2006),
planets such as Mercury, Mars, Venus or satellites around Earth
(Moon), Jupiter, Saturn, Uranus and Neptune, a lot of different data
regarding conditions of soil, atmosphere, temperatures, pressure
etc. were collected. Our goal is to make a map of as many as
possible areas of the objects in Solar system, and compare these
with extreme conditions on the Earth. Example: the temperature in
some areas on Mars, Callisto, Ganymede or Europa is similar to that
on Antartica, but it is also very important to compare other
components of extreme conditions (chemical and physical, such as
pressure or existence of O$_2$ or metan).

We intent to analyse all collected data from the NASA and the others
databases of planetary and Solar exploration, make our databases of
conditions in extreme environments, and compare these databases.
This will lead to construction of Atlas of planetary conditions
potentially suitable for life and development of web application
where conditions of an Earth extreme environment of interest could
be compared to the closest resembling counterpart in space.

During the course of analyses some COGs were found to be common to
only limited number of hyperthermophiles. For example, a {\it COG1361}, 
an S-layer protein representing a glycoprotein which is a cell wall
component of some hyperthermophiles was common to {\it P.~abyssi, 
Pyrococcus horikoshii (P.~horikashii)} and {\it M.~jannaschii}. 
This structure was previously implicated in survival of extreme 
temperatures. However, we didn't find to be present in any of 7 
hyperthermophiles analysed and therefore not presented here. Likewise, 
many other genes which were common to subgroups of the analysed 
hyperthermophiles (but not to all of them,)were not the scope of this study.

\section{6. CONCLUSIONS}

This study revealed a large number of COGs common between {\it E.~coli}
and {\it P.~abyssi} showing that such two distinct representatives of
Archaea and Bacteria have a large portion of proteins in common.

We determined 29 COGs specific to only 7 studied hyperthermophiles
and distinct from mesophiles exemplified in {\it E.~coli}. We anticipated a
smaller number of COGs to be found because the hyperthermophiles
chosen were so diverse (including one aerobe and several strict
anaerobes). Moreover, when a halophile (Halobacterium) was input in
this program, no further restriction of the common genes was
observed (data not shown).

However, since there is an extreme interest in these extremophile
organisms, the sequencing of their genomes advances with amazing
speed and the authors anticipate that the input of novel sequences
(COGs) in our program will lead to further reduction of the genes
common to hyperthermophiles.

\references

Bult, C. J., White, O., Olsen, G. J., Zhou, L., Fleischmann, R. D.,
Sutton, G. G., Blake, J. A., FitzGerald, L. M., Clayton, R. A.,
Gocayne, J. D., Kerlavage, A. R., Dougherty, B. A., Tomb, J. F.,
Adams, M. D., Reich, C. I., Overbeek, R., Kirkness, E. F.,
Weinstock, K. G., Merrick, J. M., Glodek, A., Scott, J. L.,
Geoghagen, N. S., \& Venter, J. C., 1996, Science, 273, 1058

Cavicchioli, R., Thomas, T., \& Curmi, P. M., 2000, Extremophiles, 4, 321

Cavicchioli, R., 2002, Astrobiology, 2, 281

Christner, B. C., Mosley-Thompson, E., Thompson, L. G., \& Reeve, J.
N., 2001, Environ. Microbiol., 3, 570

D'Amico, S., Claverie, P., Collins, T., Georlette, D., Gratia, E.,
Hoyoux, A., Meuwis, M. A., Feller, G., \& Gerday, C., 2002, Philos.
Trans. R. Soc. Lond. B. Biol. Sci., 357, 917

Feller, G., \& Gerday, C., 2003, Nat. Rev. Microbiol., 1, 200

Friedmann, E. I., \& Ocampo-Friedmann, R., 1995, Adv. Space. Res.,
15, 243

Goodchild, A., Raftery, M., Saunders, N. F., Guilhaus, M., \&
Cavicchioli, R., 2004, J. Proteome. Res., 3, 1164

Hearn, E. M., Dennis, J. J., Gray, M. R., \& Foght, J. M., 2003, J.
Bacteriol., 185, 6233

Jones P.A., Burton M.G., Sarkissian1 J.M., Voronkov M.A.,
Filipovi\'c M. D., 2006, MNRAS, 369, 1995

Kato, C., \& Nogi, Y., 2001, FEMS Microbiol. Ecol., 35, 223

MacDonald, A., 1987, In: M. R. Jannasch HW, Zimmerman AM (ed.), High
pressure biology, London: Academic Press, pp. 207,

Margesin, R., \& Schinner, F., 1999, Chemosphere, 38, 3463

Meadows, V., 2005, In: Modeling the diversity of extrasolar planets,
Proceedings of the International Astronomical Union, 1, 25

Meech K.J., et al., Filipovi\'c M.D., et al., Deep Impact:
Observations from a Worldwide Earth-Based Campaign, 2005, Science,
310, 265

Navarro-Gonzalez, R., Rainey, F. A., Molina, P., Bagaley, D. R.,
Hollen, B. J., de la Rosa, J., Small, A. M., Quinn, R. C.,
Grunthaner, F. J., Caceres, L., Gomez-Silva, B., \& McKay, C. P.,
2003, Science, 302, 1018

Ng, W. V., Kennedy, S. P., Mahairas, G. G., Berquist, B., Pan, M.,
Shukla, H. D., Lasky, S. R., Baliga, N. S., Thorsson, V., Sbrogna,
J., Swartzell, S., Weir, D., Hall, J., Dahl, T. A., Welti, R., Goo,
Y. A., Leithauser, B., Keller, K., Cruz, R., Danson, M. J., Hough,
D. W., Maddocks, D. G., Jablonski, P. E., Krebs, M. P., Angevine, C.
M., Dale, H., Isenbarger, T. A., Peck, R. F., Pohlschroder, M.,
Spudich, J. L., Jung, K. W., Alam, M., Freitas, T., Hou, S.,
Daniels, C. J., Dennis, P. P., Omer, A. D., Ebhardt, H., Lowe, T.
M., Liang, P., Riley, M., Hood, L., \& DasSarma, S., 2000, Proc.
Natl. Acad. Sci. USA, 97, 12176

Onstott, T. C., McGown, D., Kessler, J., Lollar, B. S., Lehmann, K.
K., \& Clifford, S. M., 2006, Astrobiology, 6, 377

Ponder, M. A., Gilmour, S. J., Bergholz, P. W., Mindock, C. A.,
Hollingsworth, R., Thomashow, M. F., \& Tiedje, J. M., 2005, FEMS
Microbiol. Ecol., 53, 103

Price, P. B., \& Sowers, T., 2004, Proc. Natl. Acad. Sci. USA,
101, 4631

Price, P. B., 2007, FEMS Microbiol. Ecol., 59, 217

Saunders, N. F., Thomas, T., Curmi, P. M., Mattick, J. S., Kuczek,
E., Slade, R., Davis, J., Franzmann, P. D., Boone, D., Rusterholtz,
K., Feldman, R., Gates, C., Bench, S., Sowers, K., Kadner, K.,
Aerts, A., Dehal, P., Detter, C., Glavina, T., Lucas, S.,
Richardson, P., Larimer, F., Hauser, L., Land, M., \& Cavicchioli,
R., 2003, Genome. Res., 13, 1580

Siddiqui, K. S., \& Cavicchioli, R., 2006, Annu. Rev. Biochem., 75,
403

Skidmore, M., Anderson, S. P., Sharp, M., Foght, J., \& Lanoil, B.
D., 2005, Appl. Environ. Microbiol., 71, 6986

Skidmore, M. L., Foght, J. M., \& Sharp, M. J., 2000, Appl. Environ.
Microbiol., 66, 3214

Tung, H. C., Bramall, N. E., \& Price, P. B., 2005, Proc. Natl.
Acad. Sci. USA, 102, 18292

\endreferences

}
\end{multicols}

\vfill\eject

{\ }



\naslov{
Primeri molekularne adaptacije na ekstremne uslove {\Z}ivotne sredine i primena
na svemirsku okolinu}


\authors{M. Filipovi{\' c}$^{1}$, S. Ognjanovi{\' c}$^{2}$ and M. Ognjanovi{\' c}$^{2}$}

\vskip3mm


\address{$^1$University of Western Sydney, Locked Bag 1797, Penrith South DC, NSW 1797, Australia}

\address{$^2$University of Minnesota, MMC 715, 1-185 Moos 420 Delaware Street SE Minneapolis, MN
55455, USA}

\vskip.7cm




\centerline{\rit }

\vskip.7cm

\begin{multicols}{2}
{


\rrm Predstav{lj}amo inicijalna istra{\zz}ivanja struktura gena odgovornih za
adaptaciju mikroskopskog {\zz}ivota u ekstremnim uslovima
na Zemlji. Ovde, preliminarno prezentujemo rezultate identifikacije
klustera orthologus groupa \textrm{(COGs)} zajedniqkih za nekoliko
hipertermofila i izuzima{nj}e onih zajedniqkih za
mezofile (ne-hipertermofile): {\it E. coli K12}, skup{lj}eni u grupu mogu{\cc}ih
proteina odgovornih za adaptaciju {\zz}ivota u ekstremnim uslovima.
Komparativna genetiqka analiza predstav{lj}a sna{\zz}no oru{dj}e u otkriva{nj}u
novih gena odgovornih za adaptaciju u ekstremnim uslovima.
Metanogeni predstav-{lj}aju jedinu grupu organizama koji mogu da
'rastu' na 0\D \textrm{C} ({\it M.~frigidum} i {\it M.~burtonii}) i 110\D \textrm{C} 
({\it M.~kandleri}). Mada, ne sve termiqke komponente adaptacije se mogu
pripisati tim novim genima, \textrm{'chaperones'} poznatiji kao toplotni udar
protein stabilizuje enzime pri pove{\cc}a{nj}u temperature. Nax ci{lj} je
korix{\cc}e{nj}e specijalno razvijenog softvera za komparativnu analizu
gena znaqajnih za adaptaciju hipertermofila. Slede{\cc}i hipertermofilski
geni su uvrxteni u softver za potrebe ove studije: {\it M.~jannaschii,
M.~kandleri, A.~fulgidus} kao i tri vrste {\it Pyrococcus}. Zajedniqki
geni, locirani su i grupisani prema {nj}ihovoj ulozi u {\cc}elijskim
procesima. Dodatni eksperimentalni podatci su neophodni za da{lj}e
izuqavanje ovih proteina.

}
\end{multicols}

\end{document}